\documentclass[aps,twocolumn,floats,superscriptaddress,prd,nofootinbib]{revtex4-1}
\usepackage{graphicx, epsfig, bm, amsmath}

\usepackage{color}
\usepackage{hyperref}
\usepackage{ wasysym }
\begin{document}


\newcommand{\lcdm}{$\Lambda$CDM}

\newcommand{\gpr}{G^{\prime}}

\newcommand{\fnl}{f_{\rm NL}}
\newcommand{\curv}{{\cal R}}

\newcommand{\bpi}{\bm{\pi}}
\newcommand{\bp}{{\bf p}}
\newcommand{\kmax}{k_{\rm max}}
\newcommand{\xmax}{x_{\rm max}}

\newcommand{\rhomax}{{\rho_{\rm max}}}
\newcommand{\rhoQ}{\rho_{21}}
\newcommand{\rhoZ}{\rho_2}
\newcommand{\rhoS}{\rho_{22}}

\newcommand{\xsc}{x_{\rm sc}}
\newcommand{\xsb}{x_{\rm sb}}
\newcommand{\xex}{x_{\rm sce}}
\newcommand{\xe}{x_{\rm e}}

\newcommand{\af}{a}
\newcommand{\epsH}{\epsilon_H}
\newcommand{\dphi}{\phi_1}
\newcommand{\bphi}{\phi_0}
\newcommand{\dotbphi}{\dot\phi_0}
\newcommand{\dotdphi}{\dot\phi_1}
\newcommand{\ep}{\epsilon_H}
\newcommand{\bk}{{\bf k}}
\newcommand{\Mpl}{M_{\rm Pl}}
\def\nn{\nonumber}

\newcommand{\bx}{{X_0}}

\def\[{\left[}
\def\]{\right]}
\def\({\left(}
\def\){\right)}

\definecolor{darkgreen}{cmyk}{0.85,0.2,1.00,0.2}
\newcommand{\peter}[1]{\textcolor{red}{[{\bf PA}: #1]}}
\newcommand{\cora}[1]{\textcolor{darkgreen}{[{\bf CD}: #1]}}
\newcommand{\wh}[1]{\textcolor{blue}{[{\bf WH}: #1]}}
\newcommand{\gB}{g_B}
\newcommand{\WP}{W}
\newcommand{\XP}{X}
\newcommand{\B}{B^{\rm Bulk}}
\newcommand{\cQ}{c_{21}}

\newcommand{\aap}{Astron. Astrophys.}


\pagestyle{plain}

\title{Bounds on non-adiabatic evolution in single-field inflation}

\author{Peter Adshead}
\affiliation{Department of Physics, University of Illinois at Urbana-Champaign, Urbana, IL 61801, U.S.A.}
\affiliation{D.A.M.T.P., Cambridge University, Cambridge, CB3 0WA, UK}
        
\author{  Wayne Hu}
\affiliation{Kavli Institute for Cosmological Physics,  Enrico Fermi Institute, University of Chicago, Chicago, IL 60637, U.S.A.}
\affiliation{Department of Astronomy \& Astrophysics, University of Chicago, Chicago IL 60637, U.S.A.}

\begin{abstract}
We examine the regime of validity of $N$-point spectra predictions of single field inflation models that invoke transient periods of non-adiabatic evolution.  Such models generate oscillatory features in these spectra spanning frequencies up to the inverse time scale of the transient feature.  To avoid strong coupling of fluctuations in these theories  this scale must be at least $\sim 10^{-2}/c_s$ of the Hubble time during inflation, where $c_s$ is the inflaton sound speed.   We show that, in such models, the signal-to-noise ratio of the bispectrum is bounded from above by that of the power spectrum, implying that searches for features due to non-adiabatic evolution are best focussed first on the latter.
\end{abstract}

\maketitle

\section{Introduction}
\label{sec:intro}

 In this work we examine the regime of validity of models of single field inflation that invoke transient periods of non-adiabatic evolution to generate features in the spectra of curvature fluctuations. We clarify the limits that can be placed on the width of a feature in the inflationary potential or sound speed, and more generally, the shortest time scale or highest energy scale that can be probed by inflationary fluctuations while the effective theory remains weakly coupled. The characterization of the region of parameter space within which perturbation theory is under control has become more important since the release of the Planck data. These data now accurately probes a range of scales large enough to encompass modes that were on the horizon and those that were above the strong-coupling scale at a single
 epoch.  Thus, the physical interpretation within the inflationary context of correlations between these scales must be treated with care.

Features in various aspects of the inflationary Lagrangian which generate a period of non-adiabatic evolution of the inflationary background are by now well established as a phenomenological way of introducing oscillatory signatures into the spectra of curvature fluctuations during inflation.  Starobinsky first noted that a potential containing a singularity in its first derivative lead to oscillations in the resulting spectrum of  curvature fluctuations \cite{Starobinsky:1992ts}. More recently Starobinsky's model has been extensively revisited \cite{Arroja:2011yu, Martin:2011sn,  Arroja:2012ae}. A step in the potential of canonical inflation 
results similarly in oscillations of the power spectrum and bispectrum as shown both numerically \cite{Adams:2001vc, Chen:2006xjb, Chen:2008wn} and  analytically \cite{Dvorkin:2009ne, Adshead:2011bw, Adshead:2011jq, Adshead:2012xz, Miranda:2012rm, Adshead:2013zfa, Miranda:2013wxa}. Oscillations in the potential lead to resonant
logarithmic oscillations in these spectra \cite{Chen:2008wn} and occur naturally in axion monodromy inflation \cite{Silverstein:2003hf,Flauger:2009ab}. Integrating out an orthogonal field in a multi-field inflationary model with a turning trajectory generates oscillatory signals \cite{Achucarro:2012sm, Achucarro:2012yr, Achucarro:2012fd, Achucarro:2013cva}. Further work on  the effects of non-adiabatic evolution during slow roll inflation includes \cite{Chen:2010bka, Chen:2011tu, Chen:2011zf}. 

Observationally, features were first invoked to explain observed glitches in the angular spectrum of temperature fluctuations in the cosmic microwave background (CMB) in the year one analysis of the Wilkinson Microwave Anisotropy Probe (WMAP) data \cite{Peiris:2003ff}. Further studies were carried out fitting for these glitches \cite{Covi:2006ci, Hamann:2007pa}. A hint of evidence for a high frequency oscillations due to a step feature in the inflationary potential was found in WMAP year 7 data by \cite{Adshead:2011jq} and at a similar significance in the Planck temperature power spectrum \cite{Ade:2013rta} with the same amplitude \cite{Miranda:2013wxa}. Evidence for oscillations due to a resonant enhancement of the fluctuations has also been reported in WMAP data by \cite{Peiris:2013opa}, however, the evidence weakened in the Planck data \cite{Ade:2013rta,Meerburg:2013dla,Easther:2013kla}. The Planck polarization power spectrum should soon provide a more definitive
empirical check of these interpretations \cite{Mortonson:2009qv}.

In this paper, we examine theoretical limits both on these interpretations and on 
the effects of non-adiabatic evolution in general by requiring that perturbation theory remains
valid throughout all aspects of their calculation. We detail how strong coupling of the fluctuations ultimately limit the shortest time scale associated with these phenomena and hence the parameters of any given model for inflaton features.  These results  parallel those found in the context of slow roll inflation \cite{Cheung:2007st, Cheung:2007sv, Leblond:2008gg, Shandera:2008ai, Baumann:2011dt, Barnaby:2011pe}, here generalized for transient violations of slow-roll. 

This work is organized as follows. In \S \ref{sec:perttheory} we study the breakdown of perturbation theory during inflation by considering the non-linearities of the equations of motion for the perturbations. In \S \ref{sec:action} we relate this breakdown to the emergence of strong coupling from the perspective of the action for the fluctuations. In \S \ref{sec:bispecandsnr} we examine bounds on the maximal bispectrum consistent with
perturbation theory, and its resulting signal to noise. In \S \ref{sec:examples} we collect some examples of non-adiabatic evolution during inflation and its impact on inflationary fluctuations. Finally we conclude in \S \ref{sec:conclusions}.

Throughout, we work in natural units where the reduced Planck mass $M_{\rm Pl} = (8\pi G_N)^{-1/2} = 1$ as well as $c = \hbar =  1$.

\section{Perturbation Breakdown}\label{sec:perttheory}

In \S\ref{sec:linear} we briefly review the calculation of $N$-point observables using
the linearized theory.   In \S\ref{sec:kinetic}, we discuss nonlinearity in the kinetic part of the field equation and the energy density in fluctuations.  We relate this nonlinearity to the
strong coupling scale of the adiabatic vacuum in \S\ref{sec:adiabatic} and discuss nonlinearity
due to non-adiabatic excitations in \S\ref{sec:excitation}.   The validity of the perturbative
calculation requires that when the $N$-point observables are formed, the modes in question
lie below all scales associated with nonlinearity. 

\subsection{Linear Perturbations}
\label{sec:linear}

We begin by 
considering the $P(X,\phi)$  theory described by the Lagrangian density
\begin{align}
\mathcal{L} = \sqrt{-g}\[\frac{R}{2} + P(X,\phi)\],
\end{align}
where the kinetic term of the inflaton field $\phi$ is
\begin{align}
X = -\frac{1}{2}g^{\mu\nu}\partial_\mu \phi \partial_\nu\phi.
\end{align}
Note that this class contains canonical single field inflation where
\begin{align}
P(X, \phi) = X - V(\phi),
\end{align} 
as well as Dirac-Born-Infeld (DBI) inflation where
\begin{align}
P(X,\phi) = \left[
1-\sqrt{1 - 2  X/T(\phi)} \right] T(\phi)- V(\phi),
\end{align}
and $T$ is the warped brane tension.
Varying the action with respect to the field yields the nonlinear field equation
\begin{align}
 \nabla^\mu(  P_{,X}\partial_\mu \phi )+ P_{,\phi} =0 .
  \label{eqn:nonlineareom}
\end{align}
This field equation enforces covariant conservation of the stress energy tensor 
$\nabla^\mu T_{\mu\nu}=0$, where
\begin{align}
T^{\mu}_{\hphantom{\mu}\nu}& = g^{\mu\alpha}P_{,X}\partial_\alpha\phi\partial_{\nu}\phi + \delta^{\mu}_{\hphantom{\mu}{\nu}}P,
\end{align}
to all orders in  field fluctuations.  Note that
\begin{align}
\rho = -T^{0}{}_{0} = P_{,X}\dot\phi^2 - P.
\label{eqn:rho}
\end{align}

To calculate the $N$-point observables, we
expand either the action or the equation of motion around a homogeneous but time-dependent background
\begin{align}
\phi({\bf x}, t) = \bphi(t) + \dphi({\bf x}, t) +  \ldots,
\end{align}
where $\bphi(t)$ solves both the homogeneous field equation
(\ref{eqn:nonlineareom}) and
the Friedmann equations
\begin{align}
H^2 = \frac{\rho}{3},  \quad
\dot H &= -H^2\epsilon_H = -P_{,X} X.
\label{eqn:friedmann}
\end{align}
We then quantize the linear fluctuations, $\dphi({\bf x}, t)$, deep within the horizon where the impact of the time dependence of the expanding background is weak and the fluctuations are assumed to obey the linearized 
field equation. 
 
Expanding the field into modes, we obtain
\begin{align}
\hat{\phi}_1({\bf x}, t) = \frac{c_s}{a \sqrt{P_{,X}}}\int \frac{d^3 k}{(2\pi)^3} \[u_{\bf k}(t)\hat a_{\bf k} e^{i {\bf k}\cdot {\bf x}}+ {\rm h.c.}\],
\end{align}
where $\hat a_{\bf k}$ and $\hat a^{\dagger}_{\bf k}$ are creation and annihilation operators satisfying the commutation relation
\begin{align}
\[\hat a_{\bf k},\hat a^{\dagger}_{\bf k'}\] = (2\pi)^3\delta({\bf k}+{\bf k'} ).
\end{align}
The modefunction, denoted here by $u$, is the canonically normalized field defined as
\begin{align}
u = \frac{ \sqrt{ P_{,X}} }{c_s}   a \dphi,
\end{align}
while the sound speed is defined as
\begin{align}
c_s^{-2} = 1+\frac{2X P_{, XX}}{P_{,X}} \Big|_{X=\bx}.
\end{align}
The adiabatic {or Bunch-Davies} vacuum state corresponds to the choice of boundary conditions such that the canonically normalized modefunction satisfies
\begin{align}
\lim_{k s \rightarrow \infty} u_k = \frac{1}{\sqrt{2 k c_s}}e^{i  k s},
\label{eqn:BD}
\end{align}
where the sound horizon is
\begin{equation}
s(t) = \int_t^0 \frac{c_s dt}{a},
\end{equation}
and $t=0$ denotes the end of inflation.
The vacuum fluctuations are then evolved with the linear equations of motion or equivalently the quadratic action, including possible
violations of the slow-roll assumptions, through sound horizon crossing  where they freeze in.  
With $y = \sqrt{2 k c_s} u_k$, the linearized field equation is
\begin{align}\label{eqn:yeqn}
        \frac{d^2y}{ds^2} + \left(k^2 - \frac{2}{s^2} \right) y = \frac{g(\ln s)}{s^2}y.
\end{align}
Here $g$ characterizes deviations from de Sitter expansion and encodes the effect of slow-roll violations
\begin{align}
        g \equiv \frac{f'' - 3 f'}{f},
\label{eqn:gfunction}
\end{align}
with  $' \equiv d/d\ln s$  and
\begin{align}\label{eqn:fdef}
	f^2 & = 8 \pi^2 \frac{\ep c_s}{H^2} \left( \frac{a H s}{c_s} \right)^2.
\end{align}
The comoving curvature power spectrum
defined as
\begin{equation}
\langle \hat\curv_{\bk} \hat\curv_{\bk'} \rangle = (2\pi)^3 \delta^3 (\bk+\bk') P_\curv(k)
\end{equation}
 is then given by
\begin{equation}
\Delta_\curv^2 \equiv \frac{k^3 P_\curv}{2\pi^2} =  \left| \frac{ xy}{f} \right|^2,
\label{eqn:power}
\end{equation}
where $x=k s$ and becomes independent of the evaluation point for $x\ll 1$.  
Higher order correlations such as the bispectrum
\begin{equation}
\langle \hat\curv_{\bk_1} \hat\curv_{\bk_2} \hat\curv_{\bk_3}\rangle = (2\pi)^3 \delta^3 (\bk_1+\bk_2+\bk_3) B_\curv(k_1,k_2,k_3)
\end{equation}
are  calculated perturbatively at tree-level in the interaction picture from the modefunction
$y$ and the higher order contributions to the Hamiltonian.

Sudden but transient violations of the slow roll assumptions, induced for example by rolling over
features in $P_{,\phi}$,  produce excitations and non-Gaussian correlations 
deep within the horizon.  The validity of perturbative calculations of these effects critically relies on the ability to linearize fluctuations around the background and compute deviations from linearity within perturbation theory. In other words, they require that the theory of the fluctuations be weakly coupled across all scales of interest, including not only the horizon scale which the curvature perturbations freeze out, but also the scale at which they were
excited as well as the scale at which the initial vacuum state before the excitation is defined.  It is therefore useful to categorize the various checks of perturbation theory in terms of the various aspects of the field equation (\ref{eqn:nonlineareom}) which can become nonlinear.   

Nonlinearity in $P_{,X}\partial_\mu\phi$ can be recast as a comparison between the kinetic energy density of the fluctuations compared with the background whereas nonlinearity in $P_{,\phi}$ is related to the sharpness of non-adiabatic features compared with the root-mean-square (rms) size of the field fluctuations.  

Field equation nonlinearity is also distinguished by whether it arises from the adiabatic vacuum fluctuations themselves or the non-adiabatic excitations.   Excitation nonlinearity places a bound on the amount of non-Gaussianity achievable within perturbation theory.  Adiabatic nonlinearity means that the still linear excitations and the resulting non-Gaussianity may not be reliably calculated since the adiabatic vacuum state itself is strongly coupled. 

\subsection{Quadratic Energy Density and Kinetic Nonlinearity}
\label{sec:kinetic}

Nonlinearity in the kinetic part of the field equation (\ref{eqn:nonlineareom}) can be quantified by comparing the
kinetic energy density in fluctuations to that of the background.
Terms that are linear in $\dphi$ vanish upon spatial or ensemble averaging the
fluctuations leaving contributions that begin at second order in perturbation theory.

It is therefore useful to introduce the quadratic Lagrangian density,
\begin{align}
\mathcal{L}_2 
= & \frac{1}{2} a^3{P_{,X}}\[\frac{{\dotdphi^{2}}}{c_s^2} -\( \frac{\partial_i \dphi}{a}\)^2 \] +\ldots,
\end{align}
Given that we are interested in subhorizon fluctuations, we ignore metric fluctuations. The $\ldots$ represent terms that depend on derivatives with respect to $\phi$; their nonlinearity will be considered in \S \ref{sec:excitation}.  Here we consider nonlinearity during adiabatic evolution, e.g.\ well after
a non-adiabatic event, where the omitted terms are negligible.

The quadratic Hamiltonian density associated with the theory gives the contribution to the
energy density contained in the field fluctuations propagating on the background. It is constructed in the usual way from the canonical momenta 
\begin{align}
\varpi({\bf x}, t) = \frac{\partial \mathcal{L}_2}{\partial \dot\dphi} = \frac{a^3P_{,X}}{c_s^2}\dot\phi_1 
\end{align}
such that
\begin{align}
\mathcal{H}_2 = \varpi \dotdphi - \mathcal{L}_2 = & \frac{1}{2}a^3{P_{,X}}\[ \frac{\dotdphi^{2}}{c_s^2} +\( \frac{\partial_i \dphi}{a}\)^2\] + \ldots
\end{align}

Now, note that in order that the theory is well defined, we require
\begin{align}
P_{,X} > 0, \quad P_{, XX} > 0.
\end{align}
The first condition ensures that the ground state energy is positive, while the second follows from the first after imposing that the fluctuations propagate subluminally. 

We can compare the quadratic Hamiltonian to the change in the total energy in the presence
of the field fluctuation.   Again since we are evaluating the energy density for subhorizon fluctuations in an adiabatic regime, we ignore
changes to the potential energy.
 Keeping quadratic contributions in the expansion of the kinetic terms in  Eq.~(\ref{eqn:rho}),
we obtain
\begin{align}\label{eqn:quadrho}
\rhoZ = & \frac{\langle\mathcal{H}_2\rangle}{a^3} + \rhoQ + \rhoS, \nonumber\\
\rhoQ 
= &    \frac{P_{, X}}{2}\(\frac{1}{c_s^2} - 1\)\[(3+2c_3)\langle{\dotdphi^2}\rangle - \frac{\langle{(\partial_i \dphi)}^2\rangle }{a^2}\], \nonumber\\
\rhoS = &(P_{,X}+2 X P_{,XX})\dot\bphi\langle \dot\phi_2 \rangle=\frac{ P_{,X}}{c_s^2}\dot\bphi  \langle\dot\phi_2\rangle ,
\end{align}
where we have introduced
\begin{align}
c_3 = \frac{X P_{, XXX}}{P_{, XX}} \Big|_{X=\bx} .
\end{align}
For example, in DBI inflation
\begin{align}
c_{3} = \frac{3}{2}\(\frac{1}{c_{s}^2} - 1\).
\label{eqn:c3DBI}
\end{align}
The term $\rhoQ$ contains  contributions  that are quadratic in $\dphi$ that are not included
in  $\mathcal{H}_2$.   These terms are associated with 
 the change in the background $P_{,X}$ due to linear and quadratic terms in the fluctuations which then change the kinetic energy density carried by  the fluctuations and background field respectively.   Note that they involve terms that appear only at the cubic
Lagrangian level.   

On the other hand, there are analogous effects from the mean of the second order
field that also renormalize the background.  If we again compare fluctuations in a slow-roll sub-horizon 
regime where $P_{,\phi}$ can be neglected, the field equation (\ref{eqn:nonlineareom}) is a conservation
law for $J_\mu = P_{,X} \partial_\mu\phi$.  Since this current is conserved exactly, it gives the second order contribution to the charge density $J_0$ from $\phi_2$ 
in terms of quadratic combinations of $\phi_1$.   For the spatially homogeneous component,
the charge is conserved and so
\begin{align}
\frac{ P_{,X}}{c_s^2} \langle \dot \phi_2\rangle =  &
- \frac{1}{2} P_{,XXX} \dot \phi_0^3\langle \dot\phi_1^2 \rangle   \\
 &
- P_{,XX} \langle  \frac{3}{2}  \dot\phi_1^2  -\frac{1}{2} \partial^i \phi_1 \partial_i \phi_1 \rangle \dot\phi_0. 
\nonumber
\end{align}
This requires that $\rhoS=-\rhoQ$ and so
\begin{equation}
\rhoZ = \frac{\langle \mathcal{H}_2\rangle}{a^3} .
\end{equation}
Thus in the energy density, the second order effects cancel the additional background
renormalizing effects of the terms quadratic in the first order terms.   

One condition for the validity of perturbation theory is that
\begin{equation}
\rho_2 < P_{,X } X  ,
\label{eqn:rho2condition}
\end{equation}
so as not to disturb the background equation of motions
(\ref{eqn:friedmann}). When this bound is violated due to excitations, one  must take into account the effects of the backreaction of the fluctuations on the background through the Friedmann equation. A related
and potentially stronger condition is that the field equation itself remains perturbative.  
Demanding that $J_0$ receives only small corrections from the terms quadratic in $\phi_1$
gives 
\begin{equation}
|\rhoQ |< P_{,X} X.
\label{eqn:rhoqcondition}
\end{equation}
By expanding $J_\mu$ to higher order in $\phi_1$, one can show that this background condition is equivalent, up to numerical factors of order unity, to the condition that the equation of motion for the fluctuations, remains perturbative.  Thus this condition is related to requiring that the fluctuations, with or without the non-adiabatic excitations, are not strongly coupled.

For models with $c_s \ll 1$, Eq.~(\ref{eqn:rhoqcondition}) is a  stronger
bound than Eq.~(\ref{eqn:rho2condition}) as well as the simple requirement that
$| \langle \dot\phi_2 \rangle| < |\dot \phi_0|$.    
For canonical fields where $P_{,X}=1$ and $c_s=1$, Eq.~(\ref{eqn:rhoqcondition}) is 
automatically satisfied since the kinetic part of the equation of motion is linear in the field.

\subsection{Adiabatic Nonlinearity}
\label{sec:adiabatic}

We can apply the linearization conditions even in the absence of a non-adiabatic
source of excitations.    For $\rho_2$ this is 
the quadratic energy density contained in the Bunch-Davies vacuum fluctuations, which of course must be canceled off by
counterterms.   For $\rhoQ$, the kinetic linearization test on the equation of motion (\ref{eqn:rhoqcondition}) determines the strong coupling
scale of the vacuum fluctuations.

With the adiabatic vacuum fluctuations from Eq.~(\ref{eqn:BD}), the
 energy to second order in
 fluctuations becomes
 \begin{align}\nn
 \rhoZ= & \frac{1}{2}P_{, X} \left\langle{ \frac{ \dot\phi_1^2 }{c_s^2} + 
 \frac{  (\partial_i \phi_1)^2}{a^{2}} }\right\rangle\nonumber\\
 =& \frac{c_s^2}{a^2}  \int \frac{d^{3}k}{(2\pi)^3}\frac{k^2}{a^2} | u_k |^2 
 =  \frac{c_s^2}{a^2}  \int \frac{d^{3}k}{(2\pi)^3}\frac{k^2}{a^2} \frac{1}{2 k c_s}.
\end{align}
Integrating up to $k_{\rm max}$ yields
\begin{align}
\rhoZ (k_{\rm max}) = & \frac{ c_s} {16\pi^2}
 \(\frac{k_{\rm max}}{a}\)^4.
 \label{eqn:rhozero}
\end{align}
 This is the familiar zero point formula for a cutoff in momentum space.    The additional
 factor of $c_s$ arises because energy is related to momentum as $\omega= k c_s/a$.
 This zero point energy is of course infinite if $\kmax \rightarrow \infty$.  
We assume that it is renormalized away with appropriate counterterms.   The direct energy comparison in
Eq.~(\ref{eqn:rho2condition}) does not place a physical bound on the adiabatic
vacuum fluctuations themselves.

 Thus the more interesting comparison is to $\rhoQ$ which tests whether the vacuum 
 fluctuations can be treated using the linearized field equation.    Here
\begin{equation}
\rhoQ = \frac{\cQ}{c_s^2} \rhoZ,
\end{equation}
where
\begin{equation}\label{eqn:c21}
\cQ =  \frac{1}{2}\(1 - c_s^2\)\[(3+2c_3)c_s^2 -1\].
\end{equation}
Note that for $c_s=1$, $\rhoQ=0$ since the field equation is linear in the absence of $P_{,\phi}$  whereas for small $c_s$, there is a $c_s^{-2}$ enhancement of $\rhoQ$.    For example in
DBI, $\cQ=1-c_s^2$ and
\begin{align}
\rhoQ   = & \frac{1-c_s^2}{c_s^2}\rhoZ = \frac{1-c_s^2}{16 \pi^2  c_s} \(\frac{ k_{\rm max}}{a}\)^4 ,\quad {\rm DBI}.
\end{align}
The condition
\begin{equation}
|\rhoQ| < P_{,X} X = H^2\epsilon_H
\end{equation}
 places a limit on the $k_{\rm max}$ for which we can reliably use linear perturbation theory. 
 It is instructive to recast this bound using the power spectrum in the adiabatic limit
 \begin{equation}
\Delta_\curv^2 = \frac{H^2}{8\pi^2 \epsilon_H c_s},
\end{equation}
and compare the wavenumber to the sound horizon
\begin{equation}
\xsc \equiv k_{\rm max} s\approx \frac{c_s k_{\rm max}}{ a H} = \frac{\omega_{\rm max}}{H}
\end{equation}
to obtain
\begin{equation}
\xsc = \left( \frac{2}{|\cQ|} \right)^{1/4} \frac{c_s}{\sqrt{\Delta_\curv}}.
\label{eqn:strongcouplinga}
 \end{equation}
For modes of higher frequency, the vacuum fluctuations no longer obey a linear equation
and hence are strongly coupled.  We call this the $c_s<1$ strong coupling scale. {Notice that one can always tune $c_3$ in (\ref{eqn:c21}) to make $c_{21}$ vanish, in which case (\ref{eqn:strongcouplinga}) diverges independently of $c_s$. At the Lagrangian level (see Sec.\ \ref{sec:sc}) this corresponds to canceling two cubic operators against each other. Since this is only possible at a fixed point in momentum space it does not apply to quartic order, and so there will still be strong coupling, however, it will arise from a different operator. For simplicity, we will assume that $c_3$ is such that $c_{21}\neq 0$ away from $c_s = 1$.}

Note that this is true even if the underlying theory is
UV complete as in the DBI case.   It represents not a breakdown of the theory itself
but rather of the calculational tools of perturbation theory. Fortunately such a situation is not catastrophic for inflationary model building. The decoupling theorem implies that as long as frequencies redshift below the strong coupling scale in the adiabatic vacuum, the low-energy physics is decoupled from the physics above this scale. This means that, provided there remains a hierarchy between the Hubble scale, the scale of the cosmological experiment where observables are determined, and the strong coupling scale, there is no catastrophic consequence to this scale.  

If on the other hand, slow-roll is interrupted by a non-adiabatic phase where the relevant
timescale is much shorter than the Hubble time, then modes with frequencies associated with this scale are
excited and can potentially be above the scale at which perturbation theory breaks down.
We turn to this case next.

\subsection{Excitation Nonlinearity}
\label{sec:excitation}

If the inflaton transits a feature in much less than an efold, its fluctuations evolve
non-adiabatically and the 
 the observable impact of nonlinearity can be greatly enhanced.
Let us consider the case where the field equation (\ref{eqn:nonlineareom}) gains
a source from a feature in $P_{,\phi}$. 
In the linearized approach these lead to large contributions to the source $g$ of modefunction excitations in Eq.~(\ref{eqn:yeqn}) as we shall see explicitly below.  
  In addition to the considerations of the previous sections for nonlinearity in $P_{,X}$, the validity of this approach requires
that $P_{,\phi}$ can be approximated in perturbation theory.

The linearization involves approximating 
\begin{equation}
P_{,\phi} - P_{,\phi}|_{\phi_0} =  P_{,\phi\phi} \phi_1 + P_{,\phi X} \dot\phi_0 \dot\phi_1 + \ldots
\label{eqn:Pphi}
\end{equation}
For a sharp feature in field space of width $\Delta\phi=d$ this approximation will
break down once the rms field fluctuation in the vacuum 
\begin{align}
\phi_{\rm rms}^2 	&  = \frac{k_{\rm max}^2 c_s }{8\pi^2 P_{,X} a^2},
\end{align}			  
 exceeds it. In this expression $k_{\rm max}$ is the maximum wave number we want to describe in our theory. This implies that perturbation theory can only be valid for all modes of interest if 
 \begin{align}
d > \frac{c_s k_{\rm max}}{a \sqrt{8\pi^2 c_s P_{,X}}}.
\end{align}
If we wish to calculate out to the $k_{\rm max}$ set by the feature width itself
\begin{align}
k_{\rm max} \approx \frac{1}{\delta s} \approx \frac{\sqrt{2 X}}{s  H d } \approx \frac{a \sqrt{2 X}}{  c_s d},
\end{align}
then  $k_{\rm max} s< \xsb$ where
\begin{align}
\xsb^2 &=  \frac{\sqrt{16 \pi^2 c_s X P_{,X}}}  {H^2} ,
 \nonumber\\
\xsb & = \frac{2^{1/4}}{\sqrt{\Delta_\curv}}.
\label{eqn:widthbound}
\end{align}
This is the so-called symmetry breaking scale \cite{Baumann:2011su, Baumann:2011ws}.   Fluctuations of higher frequency no longer experience a sharp feature due to the presence of a perturbing background of field fluctuations, regardless of the amplitude of the feature.  Above this scale it no longer makes sense to work with fluctuations about the classical background.

More specifically, if
we consider a feature in $P_{,\phi\phi}$ of amplitude $A$, then each successive
term in the expansion in Eq.~(\ref{eqn:Pphi}) will involve powers of $\phi_{\rm rms}/d$ so that
\begin{equation}
A \left( \frac{\phi_{\rm rms}}{d} \right)^n < 1.
\end{equation}
As $n \rightarrow \infty$, even $A \rightarrow 0$ becomes an uncontrolled expansion (see 
also \S \ref{sec:strongcoupling}).
Note that,  when the sound speed of the fluctuations $c_s < 1$, the scale at which the equations become non-linear due to this effect is \emph{above} the $c_s<1$ strong coupling scale in Eq.\ (\ref{eqn:strongcouplinga}) due to non-linearities in the kinetic term.  Validity of perturbation theory requires that we calculate only below the lowest of the various scales.  As we will see in the next section, violating the $c_s<1$ strong coupling bound while not violating the symmetry breaking scale still implies that the calculation of the $N$-point 
functions breaks down.

Assuming the validity of the linearized field equation, we can compute the impact of non-adiabatic feature on the modefunctions and determine the excitations on top of the
vacuum state.
 In the generalized-slow-roll (GSR) approximation \cite{Stewart:2001cd}, one first defines the solution to Eq.~(\ref{eqn:yeqn}) with
$g=0$ and Bunch-Davies initial conditions
\begin{equation}
y_0(x) = \left( 1 + {i \over x} \right) e^{i x} ,
\end{equation}
where $x= k s$ and then  replaces the RHS of Eq.~(\ref{eqn:yeqn}) with $y \rightarrow y_0$.     The solution is $y = y_0 +\delta y$ with
\begin{align}\label{eqn:firstordermode}
\delta y(x)&=  -\int_{x}^{\infty}\frac{d u }{u^2}
g(\ln  s)y_0(u)\Im[y^*_{0}(u)y_{0}(x)] .
\end{align}

These expressions simplify in the limit of subhorizon fluctuations $x \gg1$ since
$y_0= e^{ix}$ and
\begin{align}\label{eqn:firstordermode}
 y_1(x)
& = \alpha(x) e^{i x} + \beta(x) e^{-i x},
\end{align}
where
\begin{align}
\alpha(x) & =  \frac{i}{2} \int_{x}^{\infty}\frac{d u }{u^2}
g(\ln \tilde s) ,\nonumber\\
\beta(x) &= -\frac{i}{2} \int_{x}^{\infty}\frac{d u }{u^2} e^{2 i u}
g(\ln  \tilde s) ,
\label{eqn:ab}
\end{align}
and $u= k \tilde s$.
We can then iterate to obtain
\begin{align}
y_2(x) = \frac{i}{2} \int_x^\infty \frac{du }{u^2}  g [ \alpha e^{i u} + \beta e^{-i u} ] [ e^{i(x-u)}- e^{-i(x-u)}].
\end{align}
Note that only the positive frequency term of the second order field can contribute after averaging since
the negative frequency term has no unperturbed part.  Keeping only this term
\begin{align}
y_2(x) = \frac{i}{2} e^{i x} \int_x^\infty \frac{du }{u^2} g [ \alpha + \beta e^{-2i u}] + ...,
\end{align}
and using
  \begin{align}
\int_x^\infty du F(u)  \int_u^\infty {dv}F^*(v)
 + {\rm cc} = \Big| \int_x^\infty \frac{du }{u^2} F(u) \Big|^2,
\end{align}
we obtain the quadratic contributions as
\begin{align}
|y|^2 - 1 \approx  2 |\beta|^2 .
\end{align}
This is the well known Boglioubov normalization relation derived perturbatively. 
In keeping with the Boglioubov language, we retain as representative only the piece of
the modefunction change
arising from particle excitations after the non-adiabatic feature. This is the negative
frequency contribution, which reduces the total by a factor of 2.  

Note that even after $|\beta|^2$ becomes a constant in time it still depends on $k$ in a
manner that reflects the source of the excitation $g$.  
Since we are interested in the piece of the source $g$ of Eq.~(\ref{eqn:gfunction}) 
 with the highest
number of derivatives
\begin{equation}
g \approx (\ln f)'',
\end{equation}
and so it is useful to integrate by parts
\begin{equation}
\beta(x) \approx -\int_x^\infty \frac{d u}{u}   e^{2i u} (\ln f)'.
\label{eqn:betaintegral}
\end{equation}
 Suppose now that the source $(\ln f)'$ has compact support around $s_f$ with some width
  $\delta s$.  That finite width  will introduce a cutoff due to the oscillatory integrand when  $k \approx 1/\delta s$.
Thus 
\begin{align}
|\beta(x)| &\approx 
\begin{cases}
\delta\ln f,  & k \delta s\ll 1, \\
0, &  k\delta s\gg 1 .
\end{cases}
\label{eqn:betastep}
\end{align}
Given Eq.~(\ref{eqn:power}), this fractional change in the modefunction propagates  into
a change in the square root of the power spectrum as
\begin{equation}
\delta\ln \Delta_\curv \approx \delta \ln f.
\end{equation}
To maintain generality for other types of excitations where $(\ln f)'$ does not necessarily have
compact support, we use this notation from this
point on.

With this in mind the second order energy density in the excitations
\begin{align}\nn
\rho_{e2}
= &  \frac{1 }{4\pi^2  a_f^4}  \int 
 \frac{d k}{k}  k^4 |\beta|^2 ,
\end{align}
can be more instructively written by defining
\begin{equation}
 \left( \frac{\delta \Delta_\curv}{\Delta_\curv}\right)^2  \frac{1}{4 \delta s^4}  \equiv  \int \frac{d k}{k}  k^4 |\beta|^2 ,
 \label{eqn:betascaling}
\end{equation}
which can be taken as a definition of $\delta s$ and $\delta\Delta_\curv$.

  We then obtain an energy density of the same form as the zero point energy density
in Eq.~(\ref{eqn:rhozero}) at $a=a_f$
\begin{equation}
\rho_{e2} = \left( \frac{\delta \Delta_\curv}{\Delta_\curv}\right)^2  \rhoZ (1/\delta s),
\end{equation}
where the cutoff of the effective theory is replaced by the cutoff imposed by the
finite duration of the excitation.    
The energy density in excitations should not exceed the kinetic energy in the background
\begin{equation}
\rho_{e2} < \epsilon_H H^2,
\end{equation}
or
\begin{equation}
\xe = \frac{s_f}{\delta s}
< 2 ^{1/4} \frac{1} {\sqrt{  \delta \Delta_\curv }}.
\end{equation}
{Note that the bound found by requiring $k_{\rm max}s < \xe$ is always weaker than the analogous bounds derived using $ k_{\rm max}s < \xsb$ from Eq.~(\ref{eqn:widthbound}) for }
$|\delta\ln \Delta_\curv|<1$.

The analog of the $c_s<1$ strong coupling bound Eq.~(\ref{eqn:strongcouplinga}) for excitations
becomes
\begin{equation}
\rho_{e21} = \left( \frac{\delta \Delta_\curv}{\Delta_\curv}\right)^2  \rhoQ (1/\delta s) < \epsilon_H H^2,
\end{equation}
or
\begin{equation}
\xex
< \left(\frac{2}{|\cQ|}\right)^{1/4} \frac{c_s} {\sqrt{ \delta  \Delta_\curv}}.
\label{eqn:strongcouplinge}
\end{equation}
In contrast to the adiabatic $c_s<1$ strong coupling scale, this bound weakens as $\delta \Delta_\curv \rightarrow 0$.  Violation of {the bound implied by} Eq.~(\ref{eqn:strongcouplinga}) but not (\ref{eqn:strongcouplinge}) would indicate that the excitations are still linear around a vacuum state that is nonlinear.  
As such, the linearized calculation may be unreliable.  
  On the other hand, violation  of  Eq.~(\ref{eqn:strongcouplinge}) means that the excitations themselves are strongly coupled and their interactions or non-Gaussianity can no longer
  be treated perturbatively.   Thus saturation of this bound at $\delta \ln \Delta_\curv \sim 1$
   gives the maximal level of non-Gaussianity that can be achieved by non-adiabatic
   excitations.

In summary, nonlinearity in the field equation provides four characteristic spatial scales relative
to the sound horizon, or energy scale relative to the {Hubble expansion rate}.  They are 
related by 
\begin{align}
\xsb &= \frac{2^{1/4}}{\sqrt{\Delta_R}} \\
&=
\frac{ |\cQ|^{1/4} }{c_s} \xsc = \sqrt{ \frac{\delta \Delta_\curv}{\Delta_\curv}}  \xe
 = \frac{ |\cQ|^{1/4} }{c_s} \sqrt{ \frac{\delta \Delta_\curv}{\Delta_\curv}}  \xex \nonumber,
\end{align}
and can be interpreted as
$\xsc$, the $c_s<1$ strong coupling scale of vacuum fluctuations; 
$\xsb$ the symmetry breaking
scale beyond which features are blurred out by vacuum fluctuations; $\xe$ energy conservation violation scale; $\xex$ the $c_s<1$ strong coupling scale of the excitations in low sound speed models.
The validity of perturbation theory requires that any non-adiabatic feature has
a fractional temporal width $\delta s/s_f > 1/x_i$, where $x_i$ is the smallest of the four scales. We loosely refer to this smallest of the four scales as the strong coupling scale of the theory for
reasons that we make clear in the following section.
\section{Action Considerations}\label{sec:action}
In this section we relate the four scales identified through nonlinearity of the field equations 
with the corresponding scales determined by expanding the action itself.   
We review the effective field theory method of identifying 
strong coupling of the adiabatic background \cite{Cheung:2007st} and apply it to excitations.
For canonical $c_s=1$ fields, this condition involves terms that break the exact shift symmetry and
picks out the symmetry breaking scale for excitations.    In \S \ref{sec:bispectrum}, we
use the cubic action to derive a generic scaling relation for the maximal bispectrum due
to a non-adiabatic feature.

\subsection{Effective Field Theory}\label{sec:sc}

The effective field theory of inflation  \cite{Cheung:2007st} provides a useful organizational structure for considering generic higher-order terms in the inflaton action.  Given single field
inflation, there is a preferred slicing called unitary gauge  
where $\phi({\bf x},t_u)=\phi_0(t_u)$.   In this gauge the degrees of freedom are in the metric
and one can write down all possible terms
that obey unbroken spatial diffeomorphism invariance. {We are primarily interested in $P(X, \phi)$ models in this work, and so we will restrict our attention to the sector of the effective field theory which corresponds to this theory. This amounts to considering terms polynomial in $g^{00}$, while disregarding higher derivative terms such as those that arise from curvature-squared terms, as well as powers of the extrinsic curvature  \cite{Cheung:2007st}.} 
Taylor expanding the resulting function around $-1$, we obtain the effective field theory action
\begin{align}
S = \int d^4 x \sqrt{-g} \Big[ \frac{1}{2} \Mpl^2 R
 + 
\sum_{n=0}^\infty \frac{1}{n!} M_n^4(t_u) (g_u^{00}+1)^n \Big] .
\label{eqn:unitary}
\end{align}
Further demanding that its constant and
linear terms satisfy the Friedmann equations gives
\begin{align}
M_0^4 &=  -(3 H^2 + 2\dot H) \Mpl^2, \nonumber\\
M_1^4 &= \dot H \Mpl^2,
\end{align}
where we have explicitly kept $\Mpl$ to highlight the dimensions. 
We can restore temporal diffeomorphism invariance by using the St\"{u}ckel\-berg trick which
amounts to relating unitary time to an arbitrary slicing by introducing an auxiliary Goldstone field $\pi$
\begin{equation}
t_u = t + \pi(x^i, t) 
\end{equation}
and using the transformation rule for the metric 
\begin{equation}\label{metricrule}
g_u^{00} = \frac{\partial t_u }{\partial x^\mu}\frac{\partial t_u}{\partial x^\nu} g^{\mu\nu}.
\end{equation}
In the decoupling limit \cite{Cheung:2007st} we can ignore the mixing due to metric  fluctuations and set
\begin{equation}
g_u^{00} = -(1+\dot \pi)^{2}+ \frac{(\partial_i \pi)^2}{a^2}.
\end{equation}
Note that after reintroducing the Goldstone field $\pi$ and making use of Eqn. (\ref{metricrule}) 
the action at Eq.\ (\ref{eqn:unitary}) is simply a change of variable of the action of a $P(X, \phi)$ model to $\pi = \phi_1/\dot{\phi}_0$. In this case the theories can be matched by identifying
\begin{align}
M_n^4 = (-X)^n \left.\frac{\partial ^n P}{\partial X^n}\right|_{X = X_0}.
\end{align} 

Let us begin with adiabatic assumptions where $H^2$, $\dot H$ and $M_n^4$ are taken to be
approximately constant. Expanding the action in terms of $\pi$ and keeping quadratic and higher
order terms, we have
\begin{align}
S = & \int d^4 x \sqrt{-g} \sum_{m=2}^\infty \mathcal{L}_m, \\
\mathcal{L}_m =& \sum_{n=n_{m}}^m \frac{ 2^{2n-m} M_2^4 c_n \dot\pi^{2n -m} } {(m-n)! (2n-m)!} 
 \left[ \dot\pi^2-\frac{(\partial_i \pi)^2}{a^2}\right]^{m-n} ,\nonumber
\label{eqn:EFTaction}
\end{align}
where $n_{m}=m/2$ for even $m$ and $(m+1)/2$ for odd $m$. 
Here
the dimensionless $c_n$ parameters  are 
 \cite{Cheung:2007st}
\begin{align}
c_{n} = (-1)^n \frac{M_{n}^4}{M_2^4}.
\end{align}
From the quadratic Lagrangian
\begin{equation}
\mathcal{L}_2 = M_2^4 \left[ (c_1+2)  \dot\pi^2- c_1 \frac{(\partial_i \pi)^2}{a^2} \right],
\end{equation}
we can further associate 
\begin{equation}
c_s^{-2} = \frac{c_1+2}{c_1} =1-\frac{2M_2^{4}}{M_{\rm Pl}^2\dot{H}}.
\end{equation}
For DBI the $c_n$ scale as expected at low $c_s$ 
\begin{equation}
c_n = \frac{(2n-3)!!}{2^{n-2}} \left( \frac{1}{c_s^2}-1 \right)^{n-2},
\label{eqn:cnDBI}
\end{equation}
and vanish as $c_s\rightarrow 1$.

\subsection{Strong Coupling}
\label{sec:strongcoupling}

In general, one definition of strong coupling  is the lowest scale for which one of the
 operators 
\begin{align}
\left| \frac{\mathcal{L}_m}{\mathcal{L}_2}\right|_{\xsc} \sim 1,
\end{align}
for any $m>2$.   Using  Eq.~(\ref{eqn:EFTaction}), and barring any cancellation between the
$n$-indexed contributions to $\mathcal{L}_m$ we require for $c_s\ll 1$ the smallest $x_{nm}$ for which
\begin{align}
 \left( \frac{x_{nm} H \pi}{c_s^2} \right)^{m-2}  \left[
 \frac{2^{2n-m} c_n c_s^{2n-2} }{(m-n)! (2n-m)! }\right] \sim 1 \,.
 \label{eqn:strongcouplingnm}
 \end{align}
 To estimate the strong coupling scale given the power spectrum, we replace 
 \begin{align}
\pi^{2m} & \rightarrow \langle \pi^2 \rangle^{m} \nonumber\\
&  \rightarrow \lim_{x \gg 1} \left( \frac{k^3 \pi_k^2}{2\pi^2} \right)^m \sim
\left( x \frac{\Delta_\curv}{H} \right)^{2m}.
\end{align}
Note that this differs from $\langle \pi^{2m} \rangle$ by the number of contractions $2m!$
but as we shall see in the following section, the same factors come into the
calculation of  Gaussian contributions to the $N$-point functions.

Using the Stirling approximation, it is easy to show that for the DBI case the second term in 
Eq.~(\ref{eqn:strongcouplingnm}) is $[{\cal O}(1)]^{m-2}$ and hence all operators
becomes strongly coupled at the same scale  \cite{Shandera:2008ai, Baumann:2011dt}
\begin{equation}\label{eqn:dbisc}
\xsc \sim \frac{c_s}{\sqrt{\Delta_\curv}},
\end{equation}
as one might expect.
Moreover, this $\xsc$ scales in the same way as that of  Eq.\ (\ref{eqn:strongcouplinga}). This is not surprising since strong coupling is simply a breakdown of perturbativity. As we discussed above, provided there is a sufficient hierarchy  between the scales where the correlation functions are determined (e.g.\ sound horizon crossing) and this strong coupling scale there is no consequence of the existence of this scale.\footnote{It might appear
that $c_n \propto (2n-3)!!$ in DBI violates perturbative
unitarity in the S-matrix of $2\to n$ scattering as $n\to \infty$ at $x \ll \xsc$. However, one must be careful to take into account the $n$-body phase-space factors due to the final state. These can lead to additional factors of $n!$ which strongly suppress the resulting cross section
(e.g. \cite{Dicus:2004rg}).}

Thus it is the interplay of strong coupling and non-adiabatic excitations that is 
important.   While  assuming an approximate shift symmetry for the inflationary history
 is technically natural in the EFT of inflation,  it is in fact not required.   Relaxing this assumption has two effects.  For the operators included in Eq.~(\ref{eqn:EFTaction}), the $\pi$ modefunctions can gain a non-adiabatic excitation as described in \S \ref{sec:excitation}.   New operators also appear from allowing the coefficients
 in the unitary-gauge action Eq.~(\ref{eqn:unitary}) to vary.
 
We can estimate the effect of the former in a similar way to \S \ref{sec:excitation}.  Given 
a fractional change in the modefunction $\delta\ln\Delta_\curv$ that takes place across
a width in field space $\Delta\phi=d$, 
\begin{equation}
\delta\dot\pi \sim (\delta\ln\Delta_\curv) \frac{\phi_1}{d} \approx 
(\delta\ln\Delta_\curv) \frac{\dot \phi_0}{d} \pi .
\end{equation}
If we want to calculate to the highest frequency set by the width of the feature then
$x_{\rm max}  = \dot\phi_0/ d H$. Placing this scale into the comparison ${\cal L}_n/{\cal L}_2$ sets a bound on the width of the feature.   If we assume that all operators become strongly coupled at the same scale without the excitation as in the DBI scaling of
Eq.~(\ref{eqn:cnDBI}), then with the excitation we just introduce extra
factors of $\delta\ln\Delta_\curv$ from $\dot\pi \rightarrow \delta\dot \pi$. 
    If $|\delta\ln\Delta_\curv| \ll 1$, the strongest constraint is from replacing a single modefunction with the excitation
\begin{equation}
\left| \frac{{\cal L}_n}{{\cal L}_{2}}\right|_{x_{\rm max}}  \sim {\delta\ln\Delta_\curv} \left( \frac{ x_{\rm max} H \pi}{c_s^2}\right)^{n-2} \sim 1 .
\label{eqn:LratioEx}
\end{equation}
such that for sufficiently high $n$, $x_{\rm max} \rightarrow \xsc$ independently of the value of
$\delta\ln\Delta_\curv$.   This fact merely reiterates that the strong coupling bound applies
to any feature of such width or smaller.   Once the adiabatic modes become strongly coupled it does not make sense to calculate even an infinitesimal excitation on top of them.  
 For large ${\delta\ln\Delta_\curv}$ the strongest condition comes from taking all $n-2$ factors to be excited and returns
the excitation strong coupling scale $\xex$.

The second change that non-adiabatic features make is to allow new operators 
associated with expanding the $t_u$ dependent coefficients in Eq.~(\ref{eqn:unitary}) in $\pi$,
\begin{equation}
M_n^4(t_u) = M_n^4(t) + \sum_{m=0}^{\infty} \frac{M_n^{4(m)}}{m!}\pi^m,
\end{equation}
{where here and below the superscript $(m)$ on a quantity denotes $\partial_{t}^m$, e.g.\ $M_n^{4(m)} \equiv \partial^{m}_t M^4_{n}$}. For example, up to cubic order in fluctuations, the action for the effective field theory is \cite{Cheung:2007sv}
\begin{align}\nn\label{eqn:EFTnoshift}
S =&\!\! \int\!\! d^4 x\sqrt{-g}\left(- \frac{\dot{H}\Mpl^2}{c_s^2}\right)\Bigg\{ \[ \dot\pi^2 -  c_s^2\frac{(\partial_i\pi)^2}{a^2}\]  
\\ \nn& - 2\eta_H H \pi\[\dot\pi^2 -  c_s^2\frac{(\partial_i\pi)^2}{a^2}\]   - 2 H\sigma_1\pi\dot\pi^2 \\  & 
- \(1-c_s^{2}\)\dot\pi\[ \frac{(\partial_i \pi)^2}{a^2}-\(1+\frac{2}{3}c_3\)\dot\pi^2\]
+\cdots\Bigg\},
\end{align}
where $\cdots$ here refer to terms higher order in powers of $\pi, \dot\pi$ and $\partial_i \pi$ and we have dropped total time derivative terms, which yield only slow roll suppressed contributions to the bispectrum \cite{Maldacena:2002vr}. We have also dropped terms proportional to $\epsilon_H$, such as the mass term for $\pi$, consistent with taking the decoupling limit. In addition, we have introduced the slow-roll parameters
\begin{align}
\eta_H =- \frac{\ddot{H}}{2H\dot{H}}, \quad \sigma_{1} = \frac{\dot{c}_s}{H c_s}.
\end{align}
No slow roll expansion has been assumed, and at high energies, where decoupling is valid, the above action is a complete description of the interactions of $\pi$, up to cubic order. 

In order that the expansion of action remain valid, we require at least that 
\begin{align}
\left| \frac{\mathcal{L}_3}{\mathcal{L}_2} \right| \supset 2|\eta_H + \sigma_1|\pi H < 1
\label{eqn:L3etaH}
\end{align}
across all energy scales we wish to describe with our theory. Thus  we lose perturbativity in the cubic operator beyond the highest energy scale for which the above is true. Note that this was misestimated in Ref.~\cite{Adshead:2011jq}, where the perturbativity was  only required  near horizon crossing.\footnote{We note that the resulting $\xe$
bound given  in Eq.\ (\ref{eqn:featlimitpert}) was 
  already pointed out in Ref.\ \cite{Bartolo:2013exa}.  However, they do not take into account that other operators in the theory can become strongly coupled at lower scales independently of the amplitude of the feature leading to the
stronger $\xsb$ bound of Eq.\ (\ref{eqn:sbfeature}). }
From Eq.~(\ref{eqn:fdef}), we can estimate
\begin{equation}
{\rm max}({|\eta_H|,|\sigma_1|}) \sim (\ln f)'  \sim \frac{\delta \ln \Delta_\curv}{\delta \ln s}.
\end{equation}
With $\xe  =(\delta \ln s)^{-1}$,
\begin{equation}
\left| \frac{\mathcal{L}_3}{\mathcal{L}_2} \right|_{x_{\rm max}} \sim \delta \ln \Delta_\curv \left( x_{\rm max} H \pi \right)
\label{eqn:LratioV}
\end{equation}
and hence
\begin{equation}\label{eqn:featlimitpert}
x_{\rm max} < \frac{1}{\sqrt{\delta \Delta_\curv}},
\end{equation}
which is the energy bound $x_{\rm max} \sim \xe$.  If we generalize these considerations
to the interaction
\begin{equation}
\mathcal{L}_n \supset \frac{\dot H^{(n)}\Mpl^2}{n!} \pi^{n},
\label{eqn:Hdotexpansion}
\end{equation}
(see \cite{Behbahani:2011it}) then the strong coupling condition contains terms for which 
\begin{equation}
\left| \frac{\mathcal{L}_n}{\mathcal{L}_2} \right|_{x_{\rm max}} \sim \delta \ln \Delta_\curv \left( x_{\rm max} H \pi \right)^{n-2},
\label{eqn:sbfeature}
\end{equation}
where we have integrated by parts $n$ times.  Hence, in a fashion analogous to the low sound speed case of Eq.\ (\ref{eqn:LratioEx}), $x_{\rm max} \rightarrow \xsb$ as $n\to \infty$ even as $\delta \ln \Delta_\curv \rightarrow 0$.  
 Note that operators based on $\dot H(t_u)$ set the maximal scale if 
 $c_s=1$ even for models where there is no non-adiabatic features whereas those in
 Eq.~(\ref{eqn:EFTaction}) would have strong coupling scales limited only by the Planck scale.  The latter reflects the linearity of the kinetic term discussed in \S \ref{sec:kinetic}.

\section{Bispectrum and Signal to Noise}\label{sec:bispecandsnr}

Limits on the amount of nonlinearity in the field equation or the action considered in the
previous section directly translate into limits on non-Gaussianity, the observable impact 
of nonlinearity.   In this section we consider the implications for the maximal bispectrum
and the observability of features in the bispectrum vs.\ the power spectrum.

\subsection{Maximal Bispectrum}
\label{sec:bispectrum}

One advantage  of working with the action is that, in addition to giving the scales where the perturbations break down, one can also easily read off estimates of the non-Gaussianity. When the correlation functions are determined at the sound horizon, a simple estimate is found from \cite{Cheung:2007st, Baumann:2011su}
\begin{align}
\frac{k^6B_{\curv}}{(2\pi)^4\Delta^{3}_{\curv}}  \sim f_{\rm NL}\Delta_\curv \sim
 \left| \frac{\mathcal{L}_3}{\mathcal{L}_2}\right|_{x} ,
 \label{eqn:Bestimate}
\end{align}
where $x$ is the relevant scale for the formation of the bispectrum.   Thus the arguments above as to the scaling of this ratio directly give the desired result.  
For the adiabatic modes and $c_s \ll 1$, Eq.~(\ref{eqn:strongcouplingnm})  for the minimal case of $m=3, n=2$ which must be present gives
\begin{equation}
\left| \frac{\mathcal{L}_3}{\mathcal{L}_2}\right|_{x\sim 1} 
 \sim\frac{\Delta_{\curv}}{c_s^2}.
\end{equation}
For an excitation arising from a feature whose width sets $x_{\rm max}$, Eq.~(\ref{eqn:strongcouplingnm})
and (\ref{eqn:LratioV}) can be encapsulated as
\begin{equation}
\left| \frac{\mathcal{L}_3}{\mathcal{L}_2}\right|_{x_{\rm max}} 
 \sim x_{\rm max}^2 \frac{\delta\Delta_{\curv}}{c_s^2},
 \label{eqn:Bfeaturescaling}
\end{equation}
with the appropriate understanding of what generates the excitation in the $c_s \ll 1$ and
$c_s=1$ limits.
Note that the requirement that the scale set by the feature be below the strong coupling and symmetry breaking scales implies that non-Gaussianity remains small in all of the connected
 $N$-pt correlation functions
\begin{align}\label{eqn:pertdead}
\left| \frac{ \langle \curv^N \rangle_c}{\langle \curv^{2N}\rangle^{1/2}} \right|< 1
\end{align}
of which Eq.~(\ref{eqn:Bestimate}) simply provides the 3-pt version. Note that Eq.\ (\ref{eqn:pertdead}) is differs from the criteria often seen in the literature
\begin{align}\label{eqn:largenon-g}
\left|  \frac{\langle \curv^N \rangle_c}{\langle \curv^2\rangle^{N/2}} \right| < 1,
\end{align}
 by a factor of $(N!)$. A violation of the inequality in Eq.\ (\ref{eqn:pertdead}) clearly  signifies that the total $N$-point correlator has acquired strong non-Gaussianity from nonlinear terms
 whereas a violation of Eq.~(\ref{eqn:largenon-g}) does not necessarily. 
\subsection{Signal to noise ratio}

We want to determine whether or not the  signal-to-noise in the bispectrum due to some sharp non-adiabatic evolution can exceed that in the 2-pt function in a regime where
the theory is weakly coupled and all of the perturbativity criteria are met.

We can estimate the signal-to-noise ratio (SNR) in the 3-pt function in a finite volume $V$ by considering the minimum variance triangle weighting
\begin{align}
\(\frac{S}{N}\)_B^2 \approx V
 \int \frac{d^3 k_1}{(2\pi)^3}\int  \frac{d^3 k_2}{(2\pi)^3}\frac{B_\curv^2(k_1,k_2,k_3) }{6\prod_{i = 1}^{3} P_{\curv}(k_i)},
\end{align}
where ${\bf k}_1+ {\bf k}_2+{\bf k}_3=0$.   The factor of $1/6$ arises because of the 6
permutations of the triplet which are all physically identical (see, e.g.\ \cite{Hu:2000ee}, cf.\ \cite{Behbahani:2011it, Flauger:2013hra}).
After  ordering $k_1 \ge k_2 \ge k_3$ to remove the permutations, 
we can express the ratio as
\begin{align}\label{eqn:bisnr}
\(\frac{S}{N}\)_B^2 & \approx V \int \frac{d^3 k_1}{(2\pi)^3} \int^1_{1/2}\frac{dr_2}{2\pi}\int^{r_2}_{1-r_{2}} \frac{dr_3}{2\pi} r_2^4 r_3^4\\ \nn& \hskip 2.5cm\times \frac{k_{1}^{12} B_{\curv }^2(k_1, k_2, k_3)}{(2\pi^2)^3\Delta^6_{\curv}},
\end{align}
where we have defined $r_i = k_i/k_1$. We have also assumed an approximately scale invariant dimensionless power spectrum, so that $\Delta^2_\curv \approx$ const. 

We defer evaluation of the SNR for specific cases to \S \ref{sec:examples}.   Here
it is useful to extract some scaling properties.   Excitation bispectra have SNR that are peaked near equilateral triangles.   Given some limit on $x_{\rm max}$ from
the bounds of the previous section, Eq.~(\ref{eqn:Bfeaturescaling}) tells us that the equilateral
amplitude scales as
\begin{equation}
\frac{k^6 B_\curv}{(2\pi)^4  \Delta_\curv^3} \sim \frac{x_{\rm max}^2}{c_s^2} \delta \Delta_\curv.
\end{equation}
Hence ignoring factors of order unity, 
\begin{equation}
\(\frac{S}{N}\)_B^2  \sim V \int \frac{d^3 k}{(2\pi)^3}\frac{ x_{\rm max}^4}{c_s^4} (\delta\Delta_\curv)^2.
\end{equation}
This should be compared to the SNR in the power spectrum for the excitation
\begin{equation}
\(\frac{S}{N}\)_{\delta \Delta_\curv}^2 \approx V  \int \frac{d^3 k}{(2\pi)^3} ({\delta\ln \Delta_\curv})^2 .
\label{eqn:SNpower}
\end{equation}
The generic expectation for the ratio is therefore
\begin{equation}
\frac{(S/N)^2_{B}}{(S/N)^2_{\delta\Delta_\curv}} \sim \frac{x_{\rm max}^4}{c_s^4}{\Delta_\curv^2}.
\label{eqn:SNRratio}
\end{equation}
For example, given $x_{\rm max}=\xsb$ for $c_s=1$ models, this ratio can only reach
order unity.   For $c_s \ll 1$ models where the strong coupling scale enters $x_{\rm max}=\xsc$ this ratio
again can only approach unity.   Note that if either $\xe$ or $\xex$ were used, which depend on the amplitude of the excitation itself, one might get the mistaken impression that as
$\delta\ln \Delta_\curv \rightarrow 0$ that the larger SNR  is that of the
bispectrum.

A minor caveat in this analysis is that we have restricted our analysis to equilateral configurations. While this is the dominant configuration of the 3-pt function in canonical single field inflation, for models of inflation that have low sound speeds $c_s < 1$, non-adiabatic effects can induce large correlations that peak in the folded configurations due to the excitation of the non-Bunch-Davies component of the modefunctions. A detailed analysis performed for the case of resonant effects showed that one could marginally violate the bounds presented above \cite{Behbahani:2011it}, and we expect a similar result to hold more generally.

\section{Examples} \label{sec:examples}

We now consider concrete examples of models with
non-adiabatic excitations to illustrate  
the above considerations. Namely we consider models with  step and oscillatory features in \S \ref{sec:step} and \ref{sec:osc} respectively.   In both cases, the bounds on the validity of perturbation theory translate into constraints on the width of features and the non-Gaussianity they 
can produce.

\subsection{Step feature}
\label{sec:step}

A step either the potential $V(\phi)$  of a canonical $c_s=1$ inflationary model or the warp factor $T(\phi)$ in a DBI model generates a step in the function  $f$ which controls the non-adiabatic
excitation through Eq.~(\ref{eqn:gfunction}).   
A step in $f$ 
 at $s_f$ can be parameterized as
\begin{equation}
f(\ln s) = f_0 \left[ 1- \frac{{\cal C}}{4} F(\ln s-\ln s_f) \right]
\end{equation}
where
\begin{equation}
F(y) = \tanh( \pi x_d y)
\end{equation}
and $x_d$ determines the width of the step in $\ln s$.  

In the case of a potential step, the step amplitude $\Delta V/V = 2c$ is related to that in 
$f$ by \cite{Chen:2006xjb}
\begin{equation}
{\cal C} = \frac{6 c}{\epsilon_H},
\end{equation}
whereas for a warp step amplitude $\Delta T/T = 2b$  \cite{Miranda:2012rm}
\begin{equation}
{\cal C} = 2\frac{1-c_s}{1+c_s} b.
\end{equation}

In either case, for a step in field space of width
$d$, $x_d$ is approximately inverse of the fraction of the Hubble time
required for the background field
to traverse it 
\begin{equation}
x_d \approx \frac{H}{\pi d} \dot \phi_0.
\end{equation}

We can now apply the four bounds on perturbativity from the previous section.
For estimation purposes we take $\Delta_\curv = 5 \times 10^{-5}$.
The symmetry breaking bound then requires that
\begin{equation}
x_d <  \xsb \approx 170.
\end{equation}
Steps sharper than this will be blurred out by the vacuum fluctuations themselves independently of the amplitude of the step.    Likewise,  the vacuum fluctuations are
strongly coupled for some observable modes of the excitation unless
\begin{equation}
x_d < \xsc \approx 170 \frac{c_s}{|\cQ|^{1/4}}
\end{equation}
Note that for DBI $\cQ= 1-c_s^{2}$.    

For large amplitude steps the excitation nonlinearity can become the more relevant bound.
Eq.~(\ref{eqn:betaintegral}) can be approximated as
\begin{equation}
\beta \approx \frac{\cal C}{2} e^{2 i u_f} {\cal D}\left( \frac{k s_f}{x_d} \right),
\end{equation}
where the damping function
\begin{equation}\label{eqn:damping}
{\cal D}(y)=y/\sinh y.
\end{equation} 
Noting that
\begin{equation}
 \int_0^\infty y^3 {\cal D}^2(y) dy= \frac{30}{4}\zeta(5)
\end{equation}
where the Riemann $\zeta$ function $\zeta(5) \approx 1.03693$ and associating
$x_{\rm max}$ with $x_d$ and ${\cal C}$ with $\delta \ln \Delta_\curv$ we obtain
for the energy bound 
\begin{align}
x_d < \xe &= 
\left(  \frac{4}  {15 \zeta(5)} \right)^{1/4} 
 \frac{1}{\sqrt{\Delta_\curv {\cal C}} }
 \approx  \frac{100}{ \sqrt{\cal  C}} 
\end{align}
and for the excitation strong coupling bound
\begin{align}
x_d < \xex &= 
\left(  \frac{4}  {15 \zeta(5)|\cQ|}\right)^{1/4} 
 \frac{c_s}{\sqrt{\Delta_\curv {\cal C}} } \nonumber\\
 &\approx 100 \frac{c_s}{|\cQ|^{1/4} \sqrt{\cal  C}} .
\end{align}
These bounds are stronger than the symmetry breaking and adiabatic $c_s<1$ strong coupling bounds
if ${\cal C} \gtrsim 0.3$.  Note that the fractional amplitude of the oscillation in the power spectrum is ${\cal C}$ \cite{Adshead:2011jq}. 
It is interesting that the Planck CMB data favor  ${\cal C} \sim 0.23$ and $x_d \sim 10^2$ and allow larger values for each \cite{Miranda:2013wxa} (correcting \cite{Ade:2013rta}).
Thus these theoretical bounds place stronger constraints on allowed models than the
current data.

We can also estimate the SNR for the bispectrum produced by this model compared to the modification of the 2-pt function or power spectrum. For canonical single field inflation ($c_s=1$)  with a step in its potential \cite{Adshead:2011jq}
\begin{align}\nn
B_{\curv}({ k}_{1}, { k}_2, { k}_3) \approx & \frac{(2\pi)^4}{4}\frac{\Delta_{\curv}^3}{k_1^2 k_2^2 k_3^2} \[-I_0(K)+\cdots\],
\end{align}
where
\begin{align}
I_0 \sim (K s_{f})^2\frac{\mathcal{C}}{2 f_0}\mathcal{D}\left(\frac{Ks_f}{2 x_d}\right)\cos(Ks_f) 
\end{align}
and $\cdots$ refer to terms subleading in powers of $Ks_f$. Assuming a scale invariant spectrum, and approximating $\cos^2x  \approx 1/2$, we can write
\begin{align}\nn
\frac{B_{\curv }^2(k_1, k_2, k_3) }{\Delta_\curv^6}  = & \frac{\mathcal{C}^2}{4 f_0^2}\frac{(2\pi)^8}{32 }\frac{[(1+r_2+r_3)k_1 s_{f}]^4}{k_1^{12}r_2^4 r_3^4}\\ & \times \mathcal{D}^2 \left(\frac{(1+r_2+r_3)k_1s_f}{2 x_d}\right).
\end{align}
where recall $r_i=k_i/k_1$ and we have ordered $0< r_3 < r_2 < 1$. The damping function exponentially cuts off the integral in Eq.\ (\ref{eqn:bisnr}) near where
$
 k_1 \sim x_d/s_f 
$
and so we have
\begin{align}
\(\frac{S}{N}\)^2_B \sim   \frac{\mathcal{C}^2}{f_0^2}
V \int^{x_d/s_f} \frac{d^3 k_1}{(2\pi)^3} (k_1s_{f})^4.
\end{align}

We can also calculate the SNR of the modification to the 2-pt function via Eq.~(\ref{eqn:SNpower}) with 
\begin{align}
\Delta^{2}(k) \approx \Delta^2_0(k)[1+\mathcal{C}\cos(2ks_f)]
\end{align}
as
\begin{align}
\(\frac{S}{N}\)^2_{\delta\Delta_\curv}
&\sim  \frac{ \mathcal{C}^2}{2}V\int^{x_d/s_f} \frac{d^3 {k}_1}{(2\pi)^3} .
\end{align}
The ratio of these SNR gives
\begin{align}
\frac{(S/N)^2_{B}}{(S/N)^2_{\delta\Delta_\curv}}  \sim 
\frac{x_d^4}{f_0^2}
\end{align}
and note that  in order to exceed unity, one needs 
\begin{align}
x_d \gtrsim f_0^{1/2} = \frac{1}{\sqrt{\Delta_{\curv}}},
\end{align}
in  agreement with the general scaling of Eq.~(\ref{eqn:SNRratio}).   Thus, we again 
conclude  that the signal in the 3-pt function can never exceed that in the 2-pt in the canonical theory, while the theory is under control.  We emphasize here that the bound on $x_d$ is independent of the amplitude of the step and thus, regardless of the step amplitude, such a feature is better constrained by the modification to the power spectrum.

\subsection{Oscillatory features -- Monodromy}
\label{sec:osc}

Another example is that of a periodic ripple superimposed on top of an otherwise smooth slow-roll potential. This scenario was first proposed by Ref.~\cite{Chen:2008wn} before it was found to occur in axion-monodromy inflation \cite{Silverstein:2008sg, McAllister:2008hb}. In this case, inflation occurs on the potential
\begin{align}
V(\phi) = V_{\rm sr}(\phi)+\Lambda^4 \cos\(\frac{\phi}{F}\)
\end{align}
and the sound speed is canonical $c_s=1$.
Here $V_{\rm sr}(\phi)$ is a potential that supports slow roll inflation which has an additional small oscillating component. In this case $f$ oscillates as a function of $\ln s$ \cite{Flauger:2010ja, Behbahani:2011it},
\begin{align}\label{eqn:monodromyf}
f \approx 
f_0\[1+ \frac{\epsilon_{\rm osc}}{2}\cos\(\frac{\phi_k}{F}+\alpha \ln(k s)\) \],
\end{align}
where  $\phi_k$ is the value of the field when the mode with comoving wave number $k$ exits the horizon. The $\approx$ in Eq.~(\ref{eqn:monodromyf}) implies that we have expanded to first order in the small parameter $\epsilon_{\rm osc}\alpha $. We have also adopted the notation of \cite{Behbahani:2011it}, 
\begin{align}
\epsilon_{\rm osc} \equiv -6\frac{\Lambda^4 }{V_{\rm sr}' \sqrt{2\epsilon_H}}, \quad |\epsilon_{\rm osc}| < 1,
\end{align}
where
\begin{align}
\alpha = \frac{\omega}{H} = \frac{\sqrt{2\epsilon_H}}{F } \gg 1.
\end{align}
At first it may not appear obvious that this can be cast in the same way as a feature with compact support. However, as each mode redshifts, it passes through a band where it is oscillating at the same frequency as the background.  This period of resonance has compact support mode-by-mode.   Using Eq.\ (\ref{eqn:monodromyf}), Eq.~(\ref{eqn:betaintegral}) can be evaluated to a good approximation using the saddle point approximation to find \cite{Flauger:2010ja}
\begin{align}
\beta(x) \approx  \begin{cases}
\sqrt{\frac{\pi}{2}}\epsilon_{\rm osc}\alpha^{1/2}e^{-i{\phi_k}/{F}} & x<x_{\rm res}\\
0 & x > x_{\rm res}
\end{cases},
\end{align}
where
\begin{align}
x_{\rm res} = \frac{\alpha}{2}.
\end{align}
At any given time, the second order energy density stored in these excitations is then
\begin{align}
\rho_{e2} \approx \frac{\pi}{2} \frac{H^4 }{256\pi^2}\epsilon_{\rm osc}^2  \alpha^5,
\end{align}
where the $\approx$ indicates that we have assumed a hard cutoff at $k = x_{\rm res}/s$ in evaluating the integral over momenta. Identifying 
\begin{equation}\label{eqn:monopowerpert}
\delta\ln \Delta_\curv= \sqrt{\frac{\pi\alpha}{2}} \epsilon_{\rm osc},
\end{equation}
which is in fact the fractional amplitude of the oscillation in the power spectrum  \cite{Behbahani:2011it},
and $x_{\rm max}$ with $x_{\rm res} = \alpha/2$, we find
\begin{align}
x_{\rm res} <\xe = \frac{2^{1/4}}{ {\sqrt{\delta \Delta_{\curv} }}} \approx \frac{170}{\sqrt{\delta\ln \Delta_\curv}}.
\end{align}
For a small amplitude oscillation, the symmetry breaking scale is the stronger bound
\begin{align}
x_{\rm res} <\xsb \approx 170.
\end{align}
To see the origin of this bound, note that in field space, the width of the feature is $\Delta\phi \sim F$ and so the fractional transit time for the inflaton to cross it is $\Delta s/s_f \approx F/\sqrt{2 \epsilon_H} \sim (2 x_{\rm  res})^{-1}$.   Thus exceeding the symmetry breaking scale means that vacuum field fluctuations exceed the scale of the oscillatory features. Given that we have assumed $c_s=1$, the $\xsc$ and $\xex$ bounds are not relevant here. 

The oscillatory behavior of the background in axion-monodromy inflation also leads to a bispectrum \cite{McAllister:2008hb, Flauger:2010ja} 
\begin{align}
B_{\curv}(k_1, k_2, k_3) = &\frac{ (2\pi)^4 \Delta^4_{\curv}}{k_1^2 k_2^2 k_3^2}\frac{\alpha^{5/2}\epsilon_{\rm osc}\sqrt{2\pi}}{4} \sin\[\alpha \ln\({K}/{k_\star}\)\] 
\end{align}
for $\alpha\gg 1$ 
and thus, as expected
\begin{align}
\frac{k^6 B_{\curv}}{(2\pi)^4 \Delta^{3}_{\curv}} \sim x_{\rm res}^2 \delta \Delta_{\curv} ,
\end{align}
where  we have used Eq.\ (\ref{eqn:monopowerpert}).

One can also compute the SNR for the bispectrum of axion monodromy inflation. This calculation has already been performed by Ref.~\cite{Behbahani:2011it} and thus we will not repeat it here. However, note that a similar conclusion to the step feature is reached in this case. Namely, that the SNR in the bispectrum only begins  to exceed that in the power spectrum in the region where the effective theory governing the fluctuations ceases to make sense.

\section{Conclusions}\label{sec:conclusions}

In this work we have considered the bounds on non-adiabatic evolution of inflaton fluctuations during inflation. 
We have identified four frequency scales below which perturbations must lie in order to remain
perturbative: the symmetry breaking scale, the $c_s<1$ strong coupling scale, the excitation energy conservation scale, and the excitation strong coupling scale.
The first two scales place a strict lower bound on the width of any feature in the inflationary history for
which observables can be reliably computed.
This bound is independent of the amplitude of the feature, and represents the highest energy scale, or equivalently, shortest time scale at which the theory of the fluctuations is linear. Specifically, this time scale is approximately $10^{-2}/c_s$ of the Hubble time during inflation.
While, it might appear that sufficiently small perturbations should always be under control,
perturbation theory itself has broken down beyond this scale, and observables cannot be reliably computed. 

For example, a statistical description of the curvature fluctuations in terms of $N$-point functions may fail to form a perturbative series.  Although the three-point function or bispectrum may be bounded for a small
amplitude feature in the potential, higher
order operators imply an increasing amount of non-Gaussianity as $N$ increases, with
$\mathcal{L}_N/\mathcal{L}_2$ becoming order unity for a feature width at the strong coupling scale.

In a low sound speed model $\mathcal{L}_3/\mathcal{L}_2$ itself exceeds
unity as modes pass the $c_s<1$ strong coupling scale even without a feature.
Without a feature, of course, the appearance of such a scale is not a concern, provided there remains a hierarchy between this scale and the Hubble scale where all correlation functions are determined. One can then invoke the decoupling theorem to infer independence of the low energy observables from the physics above this scale. The difference in the situation at hand is that the non-adiabatic evolution may be thought of as exciting particle content at an energy scale set by the width of a feature. If this scale is above the strong coupling scale, one is in violation of the assumptions of the decoupling theorem since modes are not entering the theory in their adiabatic vacuum state.

For large amplitude features, where fractional deviations in the power spectrum approach unity, the excitations themselves violate perturbativity due to their amplitude and set the remaining two scales,
the excitation energy conservation scale and the excitation strong coupling scale.  These determine
the maximal
non-Gaussianity allowed, bounding the maximal
SNR
in the bispectrum.  Nonetheless for the detection of a non-adiabatic feature, the
{\em{relative}} SNR in the bispectrum can at most be comparable to the power spectrum
if the calculations are to remain perturbative.

While the above considerations may seem mainly of academic
interest,  current CMB data now have the dynamic range  to encompass both superhorizon and 
strongly-coupled modes at a single epoch.
Intriguingly these observations allow and even mildly favor correlated features extending to the 
strong coupling scale.  
  Convincing observational evidence for apparent features that saturate or
  exceed these strong coupling bounds  may point to new physics, or at least call into question the self-consistency of the scenarios that predict them. 

\begin{acknowledgements}

We thank Mark Wyman, Daniel Green  and especially Daniel Baumann for useful discussions and comments on a draft of this work. This work was supported in part by the Kavli Institute for Cosmological Physics at the University of Chicago through grants NSF PHY-1125897 and an endowment from the Kavli Foundation and its founder Fred Kavli. WH was additionally supported by U.S.~Dept.\ of Energy contract DE-FG02-13ER41958 and the David and Lucile Packard Foundation. P.A. gratefully acknowledge support from a Starting Grant of the European
Research Council (ERC STG grant 279617).

\end{acknowledgements}

\bibliography{FeatureBounds}

\end{document}